# Statistical study of magnetosheath jet-driven bow waves

Terry Z. Liu[1,2], Heli Hietala[3,4,5], Vassilis Angelopoulos[5], Yuri Omelchenko[6], Rami Vainio[4], and Ferdinand Plaschke[7]

[1]Cooperative Programs for the Advancement of Earth System Science, University Corporation for Atmospheric Research, Boulder, CO, USA.

[2]Geophysical Institute, University of Alaska, Fairbanks, Fairbanks, AK, USA.

[3]Department of Physics, Imperial College London, UK

[4]Department of Physics and Astronomy, University of Turku, Finland

[5]Department of Earth, Planetary, and Space Sciences, University of California, Los Angeles, USA

[6]Space Science Institute, USA

[7]Space Research Institute, Austrian Academy of Sciences, Graz, Austria

**Abstract**

When a magnetosheath jet (localized dynamic pressure enhancements) compresses ambient magnetosheath at a (relative) speed faster than the local magnetosonic speed, a bow wave or shock can form ahead of the jet. Such bow waves or shocks were recently observed to accelerate particles, thus contributing to magnetosheath heating and particle acceleration in the extended environment of Earth's bow shock. To further understand the characteristics of jet-driven bow waves, we perform a statistical study to examine which solar wind conditions favor their formation and whether it is common for them to accelerate particles. We identified 364 out of 2859 (~13%) magnetosheath jets to have a bow wave or shock ahead of them with Mach number typically larger than 1.1. We show that large solar wind plasma beta, weak interplanetary magnetic field (IMF) strength, large solar wind Alfvén Mach number, and strong solar wind dynamic pressure present favorable conditions for their formation. We also show that magnetosheath jets with bow waves or shocks are more frequently associated with higher maximum ion and electron energies than those without them, confirming that it is common for these structures to accelerate particles. In particular, magnetosheath jets with bow waves have electron energy flux enhanced on average by a factor of 2 compared to both those without bow waves and the ambient magnetosheath. Our study implies that magnetosheath jets can contribute to shock acceleration of particles especially for high Mach number shocks. Therefore, shock models should be generalized to include magnetosheath jets and concomitant particle acceleration.

**1. Introduction**

In the magnetosheath, localized fast jets with spatial scales ~1 $R_E$ are often observed (e.g., Plaschke et al., 2018 and the references therein). Such magnetosheath jets are characterized by very large dynamic pressure that are comparable to that of the solar wind. Magnetosheath jets are more likely to occur downstream of the quasi-parallel rather than the quasi-perpendicular bow shock (e.g., Plaschke et al., 2013; Vuorinen et al., 2019). One possible reason is that the surface of the quasi-parallel bow shock is rippled (e.g., Karimabadi et al, 2014; Hao et al., 2017; Gingell et al., 2017). If the solar wind crosses the bow shock where the surface is locally tilted, the downstream flow will be less thermalized and decelerated, thus forming a magnetosheath jet that is colder and faster than the ambient magnetosheath flow (e.g., Hietala et al., 2009; 2013). Sometimes, solar wind discontinuities (Archer et al., 2012) and foreshock transients (Archer et al., 2014; Omidi et al., 2016) also form magnetosheath jets.

Due to their large dynamic pressure, magnetosheath jets can disturb the magnetopause causing global perturbations in the magnetosphere-ionosphere system. Hietala et al. (2018) observed that magnetosheath jets can compress the magnetopause and trigger dayside magnetic reconnection. Archer et al. (2019) observed that magnetosheath jets can excite eigenmodes of the magnetopause surface. Such perturbations of the magnetopause can result in compressional low frequency waves in the magnetosphere, localized flow enhancements in the ionosphere, and auroral brightening (e.g., Hietala et al., 2012; Archer et al., 2013; Wang et al., 2018). Because magnetosheath jets are very common (several occurrences per hour), their geospace effect could be quite significant (Plaschke et al., 2016).

When magnetosheath jets are fast enough, they can drive a bow wave or even a shock. Using Cluster observations, Hietala et al. (2009; 2012) observed that when a supermagnetosonic (in the spacecraft frame) magnetosheath jet hit the magnetopause, it formed a secondary shock propagating sunward in the plasma frame. Simulations (e.g., Karimabadi et al., 2014) suggest that when magnetosheath jets compress the ambient plasma, a compressional bow wave or a secondary shock can form ahead of them. Liu et al. (2019a) observed such a bow wave driven by a fast magnetosheath jet and demonstrated that the bow wave can accelerate ions to tens of keV. In the paper accompanying this work, Liu et al. (2019b submitted to JGR) show that the magnetosheath jet-driven bow waves can also accelerate electrons through the shock drift/fast Fermi acceleration.

The particle acceleration by magnetosheath jet-driven bow waves could play an important role in the parent bow shock acceleration. The most widely employed shock acceleration mechanism is diffusive shock acceleration (e.g., Krymskyy, 1977; Axford et al. 1977; Bell 1978; Blandford & Ostriker, 1978; Lee et al., 2012) whereby particles are scattered back and forth between the converging upstream and downstream flow gaining energy through Fermi acceleration. While particles are scattered in the downstream region, magnetosheath jet-driven bow waves/shocks can provide additional acceleration and thus increase the efficiency of diffusive shock acceleration. Therefore, it is necessary to apply a statistical study to further investigate the properties of magnetosheath jet-driven bow waves/shocks such as their favorable solar wind conditions and whether it is common for them to accelerate particles.

In Section 2, we will describe how we selected events with the help of an example. In Section 3, we will show our statistical results on the favorable solar wind conditions and particle energization. We will conclude and discuss our results in Section 4.

## 2. Data and Methods

We used data from the THEMIS mission probes (Angelopoulos, 2008). We analyzed plasma data from the electrostatic analyzer (ESA) (McFadden et al., 2008) and the solid state telescope (SST) (Angelopoulos, 2008) and magnetic field data from the fluxgate magnetometer (Auster et al., 2008). We used OMNI data to infer the solar wind parameters.

We selected magnetosheath jets that have a bow wave or shock by searching the event list reported by Plaschke et al. (2013). As an example, here we use a magnetosheath jet (Figure 1) reported in the accompanying paper (Liu et al., 2019b submitted to JGR), to help explain how we automatically select events. We define $t_0$ as the time of the maximum dynamic pressure within magnetosheath jets (vertical solid lines in Figure 1). In order for a magnetosheath jet to supermagnetosonically compress the ambient magnetosheath plasma, the jet should be at least supermagnetosonic in the spacecraft frame. Thus, we calculate the maximum speed $V_m$ during $t_0 \pm 5s$ (yellow region in Figure 1) and require that $V_m$ is larger than the local fast wave speed (averaged within $t_0 \pm 10s$). In this case, the maximum dynamic pressure is dominated by $V_m$. Thus $t_0$ also corresponds to a time when the velocity-based magnetosheath-to-jet transition ends. Because bow waves/shocks have density and field strength enhancements and velocity deflection, we calculate the maximum value of density $\rho_m$, field strength $B_m$, and velocity vector $\boldsymbol{V}_m$ during $t_0 \pm 5s$. We then calculate the average density $\rho_{bg}$, field strength $B_{bg}$, and velocity vector $\boldsymbol{V}_{bg}$ in the background time interval defined as $[t_0 - 120s, t_0 - 20s]$ (blue region in Figure 1). We require: $\rho_m/\rho_{bg} > 1.2$; $B_m/B_{bg} > 1.2$; $|V_m - V_{bg}| > 0.4|V_{bg}|$. The reason we chose the background time interval 20s before $t_0$ is to ensure the parameter transition is (relatively) sharp. By applying our criteria, we selected supermagnetosonic magnetosheath jets that have sharp velocity transition with density and field strength enhancements at their leading edge. We then checked each event and removed those surrounded with large amplitude fast-mode fluctuations in the turbulent magnetosheath, because such events could just be one of the fluctuations. Finally, we obtained 364 events out of 2859 (~13%) that have a bow wave/shock.

Some of these bow waves may have steepened into shocks. For the given example event (Figure 1), Liu et al. (2019b submitted to JGR) calculated the normal, [-0.86, -0.48, 0.08] with uncertainty ~5.8°, using the mixed-mode coplanarity method (Schwartz, 1998) and the normal speed in the spacecraft frame, 521±69 km/s earthward, using conservation of mass flux (Schwartz, 1998). This results in a Mach number of 1.4±0.2. Across the transition layer, the ion spectrum broadens (Figure 1d, e) and the electron energy becomes locally enhanced (Figure 1g), indicating shock heating. Using similar methods, we randomly selected 37 out of 364 events to calculate their shock parameters. Our calculation shows that 36 out of 37 (~97%) events have Mach number larger than 1 and 34 events (~92%) larger than 1.1 (see Table S1), indicating that many of them may have steepened or will steepen into shocks. Here we do not distinguish between bow waves and shocks because they could represent the same structure in different stages, and both can accelerate particles.

To determine the statistical characteristics of jet-driven bow waves/shocks, we separated 2859 magnetosheath jets into two groups, with 2495 magnetosheath jets without a bow wave/shock (non-shock events) and 364 magnetosheath jets with a bow wave/shock (shock-like events). For each magnetosheath jet the corresponding solar wind parameters are obtained from the OMNI database for time $t_0$ (averaged over 5 minutes, which covers the time delay from the bow shock to jets, that is typically 1-2 minutes; Plaschke et al., 2013). In each group, we calculate the number of events $N_i$ within a certain solar wind parameter range $[\alpha_i, \alpha_i + \Delta\alpha]$. From $P_i = N_i/N$, we obtain the probability distribution as a function of the solar wind parameter, $P(\alpha)$, with a relative error $\sqrt{(1 - \frac{N_i}{N})/N_i}$, where N=2495 or 364. From the time spent by the THEMIS spacecraft in the magnetosheath during the event list (2008-2011; Plaschke et al., 2013), we obtain the amount of time $\Delta t_i$ when the solar wind parameter is within $[\alpha_i, \alpha_i + \Delta\alpha]$. By calculating $N_i/\Delta t_i$, we thus obtain the number of events per hour in the magnetosheath within $[\alpha_i, \alpha_i + \Delta\alpha]$. Then we can determine whether the occurrence of events depends on the solar wind parameter $\alpha$. To examine whether such dependencies show differences between the two groups, we also calculate the ratio between their probability distributions and the relative error (the sum of two probability distribution relative errors).

To determine whether there is acceleration/heating associated with magnetosheath jets, we calculate the highest energy channel that has energy flux larger than the instrumental noise level (black lines in Figures 1d-g). In the example event, we see that there is moderate increase of the black lines in Figures 1d, g, indicating ion and electron energy enhancements around the bow wave. For each magnetosheath jet, we calculate the maximum values of the black lines during $[t_s - 15s, t_0 + 20s]$ (maximum energy), where $t_s$ is the start time of magnetosheath jets when the dynamic pressure exceeds a quarter of the solar wind dynamic pressure (two vertical dotted lines and labeled as $\Delta t$ in Figure 1). Then we plot the probability distributions of particle energies for the two groups and calculate their ratio to determine whether there are differences.

To further confirm that there is particle acceleration/heating, we compare the ion and electron energy flux around jets with those in the ambient magnetosheath. For each event, we obtain a particle energy spectrum at the time of the maximum energy (maximum value of the black lines in Figures 1d, g) during $\Delta t$ and an averaged spectrum in the ambient magnetosheath during $[t_0 - 400s, t_s - 15s]$. Here we choose a longer time interval than the blue region in Figure 1, because Liu et al. (2019a) and Liu et al. (2019b submitted to JGR) show that the bow wave-accelerated particles can be reflected upstream, adulterating our background energy flux measurements – a longer interval averaging can reduce this effect. For each event at each energy channel, we obtain averaged background energy flux and the energy flux corresponding to the maximum energy during $\Delta t$. At each energy channel, we calculate the value corresponding to the lower quartile (12.5%), median, and upper quartile (87.5%) for the background energy flux and the energy flux of maximum energy and examine whether there are differences between them.

## 3. Results

### 3.1. Solar wind conditions

Figures 2 – 4 show the number of events per hour in the magnetosheath as a function of various solar wind parameters. Consistent with a previous statistical study for all magnetosheath jets (Plaschke et al., 2013), non-shock events do not show a dependence on dynamic pressure with maximum occurrence at ~2.5-3.0 nPa (Figure 2a). Although shock-like events (Figure 2b) have similar maximum occurrence, they show additional tendency for stronger solar wind dynamic pressure to favor their occurrence. To emphasize this effect, Figure 2c shows the probability distribution ratio of shock-like events to non-shock events. Larger ratios mean that shock-like events are more likely to occur than non-shock ones at a certain solar wind parameter range. We see that these ratios increase with the solar wind dynamic pressure. Thus, larger solar wind dynamic pressure creates a favorable condition for shock-like events.

Next, we determine whether this is caused by the solar wind density (Figures 2d – f) or the solar wind speed (Figures 2g – i). With regard to the solar wind density, Figures 2d and 2e show similar trends, and their probability distribution ratio does not show clear dependences (Figure 2f). As for the solar wind speed, Figures 2g and 2h are also very similar showing a positive tendency with the solar wind speed, but their probability distribution ratio does not show any dependences either (Figure 2i). Therefore, neither the increased solar wind density nor the increased solar wind speed generates favorable conditions alone.

Next, we look at the solar wind plasma beta (Figures 3a – c). Figure 3a does not show any trends, but Figure 3b shows a strong tendency for high plasma beta to favor the occurrence of shock-like events. As a result, their probability distribution ratio (Figure 3c) shows a clear trend that the larger solar wind plasma beta has a higher probability to form shock-like events than non-shock events. We then check whether this trend is caused by the solar wind magnetic pressure (Figures 3d – f) or thermal pressure (Figures 3g – i). Figure 3f shows a clear trend for shock-like events to occur preferentially at lower magnetic pressure, whereas Figure 3i does not reveal a clear trend with the solar wind thermal pressure. Therefore, the solar wind magnetic pressure seems to be the main controlling factor in defining the dependence of shock-like event occurrence on the solar wind plasma beta.

With regard to the solar wind Alfvén Mach number (Figure 4a – c), no trends are seen for non-shock events (Figure 4a) whereas the occurrence of shock-like events increases with the Alfvén Mach number (Figure 4b). As a result, their probability distribution ratio clearly increases with Alfvén Mach number (Figure 4c). Note that the ratio of the Alfvén Mach number to the square root of plasma beta is proportional to the ratio of the solar wind speed to the solar wind thermal speed. Since the faster solar wind is hotter, the Alfvén Mach number is proportional to the plasma beta (see Figure S1a). Thus, Figure 3c and Figure 4c are related because of this solar wind speed-temperature correlation. We will discuss how the Alfvén Mach number and plasma beta could affect the formation of shock-like events in Section 4.

As for the IMF cone angle, Figures 4d and 4e show negative trends with the IMF cone angle, consistent with the typical property of subsolar magnetosheath jets to occur more frequently during intervals of small cone angle (Plaschke et al., 2013). Their probability distribution ratio also shows weak negative trends (Figure 4f). Thus, smaller IMF cone angles may be a favorable condition for shock-like events. The IMF cone angle could be considered as a proxy for the angle between the IMF and the bow shock normal, $\theta_{BN}$, for subsolar jets. However, it is very difficult to trace the magnetosheath jets backward to the bow shock to calculate local $\theta_{BN}$. Whether the local $\theta_{BN}$ plays any role requires more comprehensive studies in the future.

Next, we compare the jets' position relative to the bow shock (1.0 in Figures 4g-i) and the magnetopause (0.0), calculated from the radial distance between the spacecraft position and the model magnetopause (Shue et al., 1998) and normalized by the radial distance between the model bow shock (Merka et al., 2005) and the model magnetopause. Figure 4g shows that there is a higher occurrence of non-shock events closer to the bow shock, consistent with the statistical study by Plaschke et al. (2013). As for shock-like events, Figure 4h shows that for relative positions less than 0.75, the trend is similar to Figure 4g, but for greater than 0.75 (near the bow shock) their occurrence decreases. Therefore, their probability distribution ratio (Figure 4i) shows that shock-like events are much less probable near the bow shock than non-shock events. There are multiple possible explanations. One is that the bow wave/shock needs time to form while the jet propagates away from the bow shock. Another is that deeper in the magnetosheath the flow direction is more along the magnetopause surface, whereas the selection of magnetosheath jets requires that the dynamic pressure in the anti-sunward direction exceeds half of the solar wind dynamic pressure everywhere in the magnetosheath (Plaschke et al., 2013). As a result, deeper in the magnetosheath the velocity difference between jets and the ambient magnetosheath could be larger than that near the bow shock.

In summary, as expected, the occurrence rates of non-shock events as a function of solar wind parameters are consistent with the previous statistical study for all the magnetosheath jets (Plaschke et al., 2013). However, shock-like events show different properties indicating that larger solar wind plasma beta (due to lower magnetic field strength), larger Aflvén Mach number, and larger solar wind dynamic pressure present favorable conditions for their formation. Small IMF cone angles may also play a role. Shock-like events are less likely to occur near the bow shock compared to non-shock events. Hybrid simulations could test these formation conditions in the future.

### 3.2. Particle energies

To investigate whether there is particle acceleration/heating associated with jet-driven shocks, we plot the probability distribution of the maximum ion energy and maximum electron energy for the two groups (Figures 5 and 6). We first compare the ion energies. Because 73% (265/364) of shock-like events have SST data (from 30 keV to 700 keV) available, we only include events that have SST data for both groups (1619/2495 ~ 65% of non-shock events). From the probability distributions of the maximum ion energy (Figure 5a, b), we see that both the non-shock events and shock-like events have the maximum probability at ~60-100 keV. This energy range is typical for the magnetosheath ions (e.g., Figure 1d, black line). Above 100 keV, shock-like events have a higher probability than non-shock events. If we calculate the ratio of two probability distributions (Figure 5c), we see more clearly that the shock-like events are more likely associated with higher maximum ion energies than non-shock events.

Before we conclude that this is due to ion acceleration/heating at the magnetosheath jet-driven bow waves/shocks, we need to exclude the possibility that this result is not due to the formation conditions. If shock-like events are biased to occur more often at higher magnetosheath ion temperatures (either due to their formation process or due to the statistical fluctuations in our dataset), there will also be higher ion maximum energies associated with shock-like events. We thus plot the probability distributions as a function of the ambient magnetosheath ion temperature (average value in $[t_0 - 120s, t_0 - 20s]$) for the two groups as well as their ratio (Figures 5d-f). The latter, in Figure 5f, does not reveal any bias. In fact, the magnetosheath ion temperature is proportional to the solar wind speed (see Figure S1b) and we have shown that the solar wind speed does not affect the occurrence of shock-like events. Thus, we confirm that the positive tendency in Figure 5c is not due to the formation conditions but caused by additional acceleration/heating by jet-driven bow waves/shocks.

Next, we compare the electron energies. Only around 50% of shock-like events have electron SST data available. To avoid biases, we plot probability distributions of events with both ESA and SST data and with only ESA data separately. Figures 6a-c show the probability distributions as a function of the maximum electron energy measured by both ESA and SST. We see that in both ESA and SST energy ranges shock-like events are more likely associated with higher electron energies than non-shock events. Figures 6d-f show probability distributions measured only by ESA. There are clear trends that shock-like events have a higher probability to be associated with higher electron energies than non-shock events. Similar to the ions, we also plot the probability distributions of the ambient magnetosheath electron temperature to check for a possible formation condition effect: no preferences for shock-like events are seen (Figures 6g-i). Thus, our results indicate that shock-like events are more likely associated with electron heating/acceleration than non-shock events.

To further examine whether there are energy increases, we compare the electron energy flux in the background and around the jets. Figure 7a is the superposed electron energy flux spectra (measured by the ESA) of shock-like events (red), non-shock events (blue), and the ambient magnetosheath (black), respectively. We see that the blue lines and black lines almost overlap, but the red lines show larger energy flux than the blue and black lines above ~100 eV, indicating that shock-like events have enhanced electron energy flux. To see this more clearly, we calculate the ratio of lower-quartile, median, and upper-quartile values between non-shock events and the ambient magnetosheath (Figure 7b) and between shock-like events and the ambient magnetosheath (Figure 7c), respectively. We see that for non-shock events the ratio is ~1 at all energy channels meaning that magnetosheath jets without a bow wave/shock do not change electron energy spectra in the ambient magnetosheath. As for shock-like events, we see that above 100 eV the ratios are ~2. This indicates that when there is a bow wave/shock, magnetosheath jets increase the electron energy flux by a factor of 2 on average at energies above 100 eV. This result is consistent with a multi-case study in the accompanying paper (Liu et al., 2019b submitted to JGR) demonstrating that the bow wave/shock can enhance the electron energy flux at energies above 100-200 eV. Our statistical results further confirm that electron acceleration/heating at these energies is common and systematic.

We also find that magnetosheath jets without a bow wave/shock do not show clear ion energy flux increases (Figure S2). However, when there is a bow wave/shock, the ion energy flux increases at energy channels above tens of keV, consistent with Figure 5c. We do not show it here as the ion distributions cannot be simply described by omni-directional spectra. This is because ions are typically anisotropic, sometimes with multiple components (e.g., Liu et al., 2019a), and the ion bulk velocity is comparable to the thermal velocity.

## 4. Conclusions and Discussion

After examining 2859 supermagnetosonic magnetosheath jets with increases in field strength and density and large velocity deflection at the leading edge, we identified 364 jets (~13%) that have a bow wave/shock ahead of them. By examining probability distributions of various plasma parameters as a function of various solar wind parameters and comparing those amongst various databases (with/without bow waves or shocks and for the ambient plasma), we conclude that large plasma beta (or low IMF field strength), large Alfvén Mach number, large solar wind dynamic pressure, and likely also small IMF cone angles are favorable solar wind conditions for shock-like events. We also show that shock-like events are less likely to occur near the bow shock, indicating either temporal or spatial requirement of bow wave/shock formation. By comparing the probability distributions of the maximum ion and electron energies, we show that irrespective of the ambient magnetosheath temperature, shock-like events are more likely associated with higher ion energy and electron energy than non-shock events. By comparing electron energy spectra in the ambient magnetosheath and around the jets, we show that shock-like events have higher energy flux (by a factor of 2 on average) above ~100 eV than the background and non-shock events. This indicates that particle acceleration/heating at magnetosheath jet-driven bow waves/shocks is common.

Such particle acceleration/heating at jet-driven bow waves/shocks could contribute to particle acceleration at parent bow shock (Figure 8). Here we estimate their contribution to commonly employed shock acceleration model, diffusive shock acceleration (e.g., Krymskyy, 1977; Axford et al. 1977; Bell 1978; Blandford & Ostriker, 1978; Lee et al., 2012), in the environment of Earth's bow shock. Depending on solar wind conditions (Figures 2-4), the occurrence of jet-driven bow waves/shocks ranged from 0.1-1.0 per hour. Here we use 0.2 per hour as an average value. Based on case studies on ion acceleration by Liu et al. (2019a) and electron acceleration in the accompanying paper (Liu et al., 2019b, submitted to JGR), when ions and electrons are reflected by a bow wave, they can gain twice of the de Hoffmann-Teller velocity (typically thousands of km/s) through shock drift acceleration. For bow waves with magnetic compression ratio of 2 (loss cone angle of 45°), 50% of incoming suprathermal particles can be reflected and accelerated. We estimate bow waves can exist at least 1-2 minutes to accelerate particles. Therefore, the average velocity gained by particles from bow waves is 50%×0.2×1.5/60×thousands of km/s ~ a few to tens of km/s. For diffusive shock acceleration, each time particles bounce across the bow shock, they gain velocity comparable to the velocity difference between the solar wind and magnetosheath (several hundred km/s) (e.g., Drury, 1983). By including jet-driven bow waves/shocks, each time particles enter the magnetosheath, they gain additional a few to tens of km/s. In other words, jet-driven bow waves/shocks can contribute additional a few to ten percent acceleration, i.e., first order modification to diffusive shock acceleration model.

Magnetosheath jets, in principle, should also exist in other shock systems. Our statistical study shows that higher Alfvén Mach number can results in more magnetosheath jet-driven bow waves/shocks (Figures 4b). Certain astrophysical shocks ($M_A$>100; e.g., Treumann, 2009) and other planetary shocks such as Saturn's bow shock ($M_A$~20; e.g., Masters et al., 2016) can have much higher Alfvén Mach numbers than Earth's bow shock ($M_A$~3-10; e.g., Balogh et al., 2005). Therefore, magnetosheath jet-driven bow waves/shocks may be even more common at other space plasma shocks that have stronger $M_A$ and could play a more important role there.

Magnetosheath jets are nonlinear structures downstream of shocks. Upstream of shocks, there are also many nonlinear structures called foreshock transients (Eastwood et al., 2005). Foreshock transients that expand fast enough can also form secondary shocks (see sketch in Figure 8). Liu et al. (2016) and Liu et al. (2019c) found that such secondary shocks can accelerate ambient solar wind and foreshock particles similar to magnetosheath jet-driven bow waves which accelerate ambient magnetosheath particles (Liu et al., 2019a; 2019b submitted to JGR). A recent statistical study by Liu et al. (2017) also demonstrated that it is common for foreshock transients to accelerate ions and electrons and possibly provide another first-order modification to shock acceleration models. Therefore, nonlinear structures both downstream and upstream of shocks could provide additional acceleration and thus play a role in shock acceleration by increasing the parent shock acceleration efficiency and providing a seed population for further acceleration. The relevant shock environment comprises not just the shock itself, but also numerous surrounding nonlinear structures with secondary bow waves/shocks, which should be included in future generalized shock models.

Here we discuss why shock-like events more likely occur at large solar wind beta, Alfvén Mach number, and solar wind dynamic pressure. The dynamic pressure of magnetosheath jets is proportional to the solar wind dynamic pressure due to the selection criteria/definition (Plaschke et al., 2013). Because the ambient magnetosheath dynamic pressure is also proportional to the solar wind dynamic pressure, higher solar wind dynamic pressure will result in larger dynamic pressure differences between magnetosheath jets and the ambient magnetosheath. A larger dynamic pressure difference means that there is more free kinetic energy to generate bow waves/shocks. Additionally, because the faster solar wind is hotter (e.g., Elliott et al., 2010), both the solar wind beta and Alfvén Mach number are proportional to the ratio of solar wind dynamic pressure to the magnetic pressure. Therefore, larger solar wind beta and Alfvén Mach number mean that there are larger kinetic energies against smaller magnetic pressure, which makes it easier for magnetosheath jets to drive bow waves/shocks. Another possibility is that higher Alfvén Mach number could cause the bow shock surface to be wavier. When the solar wind crosses more inclined portions of the bow shock surface, more kinetic energy remains to drive a bow wave. In the future, more comprehensive studies are needed to examine these possibilities.

One could argue that higher solar wind speed can cause a larger velocity difference between the jets and ambient magnetosheath and therefore should be a favorable condition. However, as the magnetosheath ion temperature is also proportional to the solar wind speed, higher solar wind speed results in larger magnetosheath fast wave speed, making it more difficult for bow waves/shocks to form. These two effects would counter each other, and thus may diminish any dependence of shock-like event occurrence on solar wind speed.

As our selection criteria, we calculate plasma parameters within $t_0 \pm 5$s, based on the assumption that $t_0$ is the time where the parameter transition ends. Here we use a very short time interval (corresponding to three data points of plasma measurements), because we want to ensure that the enhancements of magnetic field strength and density occur simultaneously, i.e., represent a fast mode compression. Such a short time interval, however, underestimates the maximum field strength and density enhancements (e.g., Figure 1) and possibly overlooks some shock-like events. For example, if we increase the time interval to 14s, the event list increases by 10%. Additionally, to ensure that the parameter transition ahead of jets is relatively sharp, we calculate the background parameters 20s before $t_0$ corresponding to a distance around several thousand km. However, sometimes the time interval of magnetosheath jets lasts several minutes and strong fluctuations inside them can violate the assumption that $t_0$ corresponds to a time when the parameter transition ends. In other words, some bow waves could be one minute before $t_0$ and thus could get excluded because the background parameters are contaminated. For example, 155 events (out of 2859; 5%) have the jet start time $t_s$ 60s before $t_0$. If we could set up better criteria to select shock-like events, statistically significant differences from non-shock events may be even more noticeable.

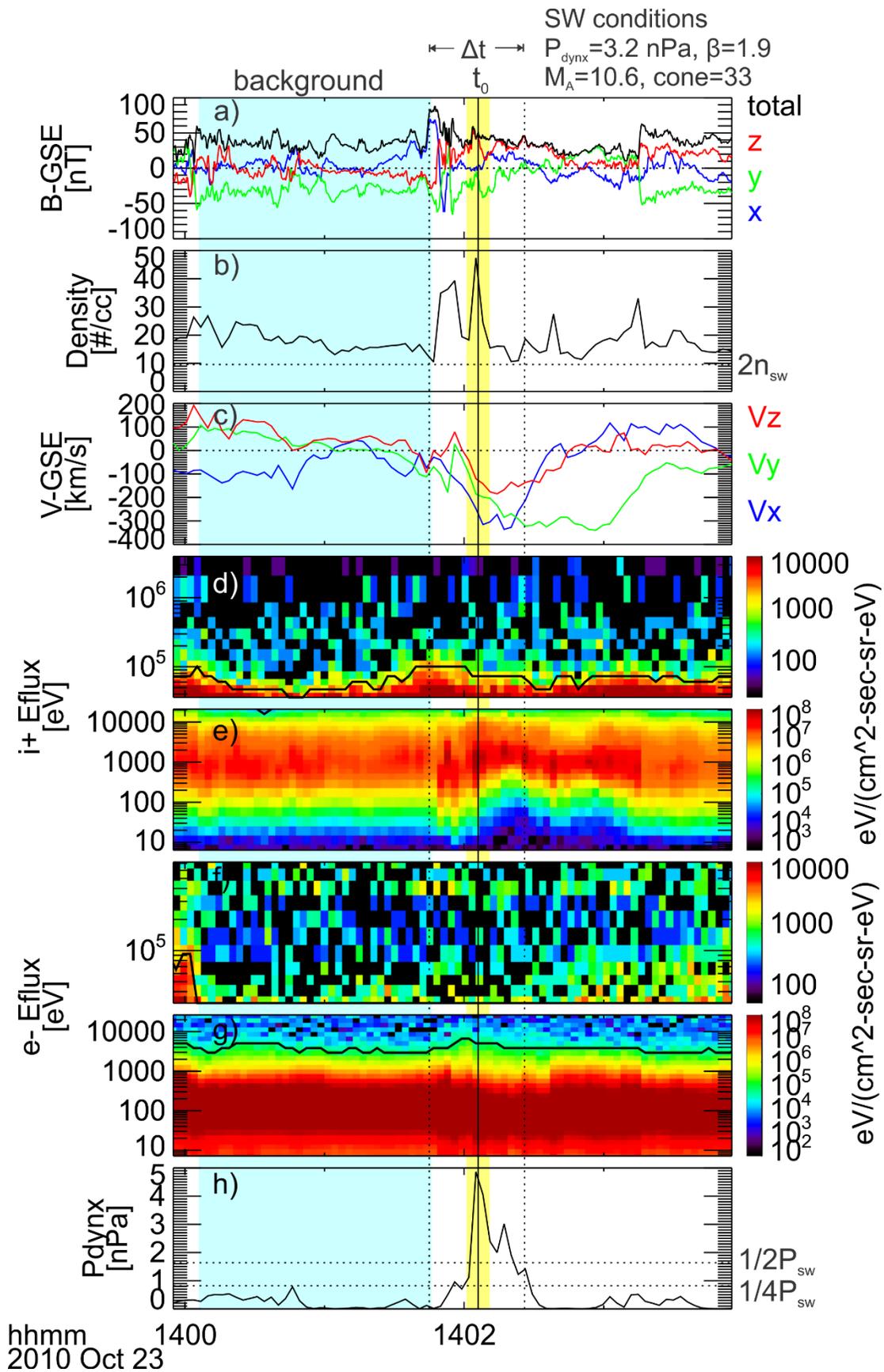

**Figure 1.** An example of magnetosheath jet-driven shock observed by THEMIS. From top to bottom: (a) magnetic field in GSE (XYZ, total in blue, green, red, and black respectively); (b) ion density (dotted line indicates two times the solar wind density); (c) ion bulk velocity in GSE (XYZ in blue, green, and red respectively); (d) ion energy flux spectrum from 30 keV to 700 keV; (e) ion energy flux spectrum from 7 eV to 25 keV; (f) electron energy flux spectrum from 30 keV to 700 keV; (g) electron energy flux spectrum from 7 eV to 25 keV; (h) dynamic pressure in GSE-X (dotted lines indicate 1/2 and 1/4 of solar wind dynamic pressure respectively). Black lines in ion and electron spectra indicate the highest energy channel that have energy flux larger than the instrumental noise level. The blue and yellow regions indicate the time intervals used to calculate the ambient magnetosheath parameters and the jet parameters, respectively.

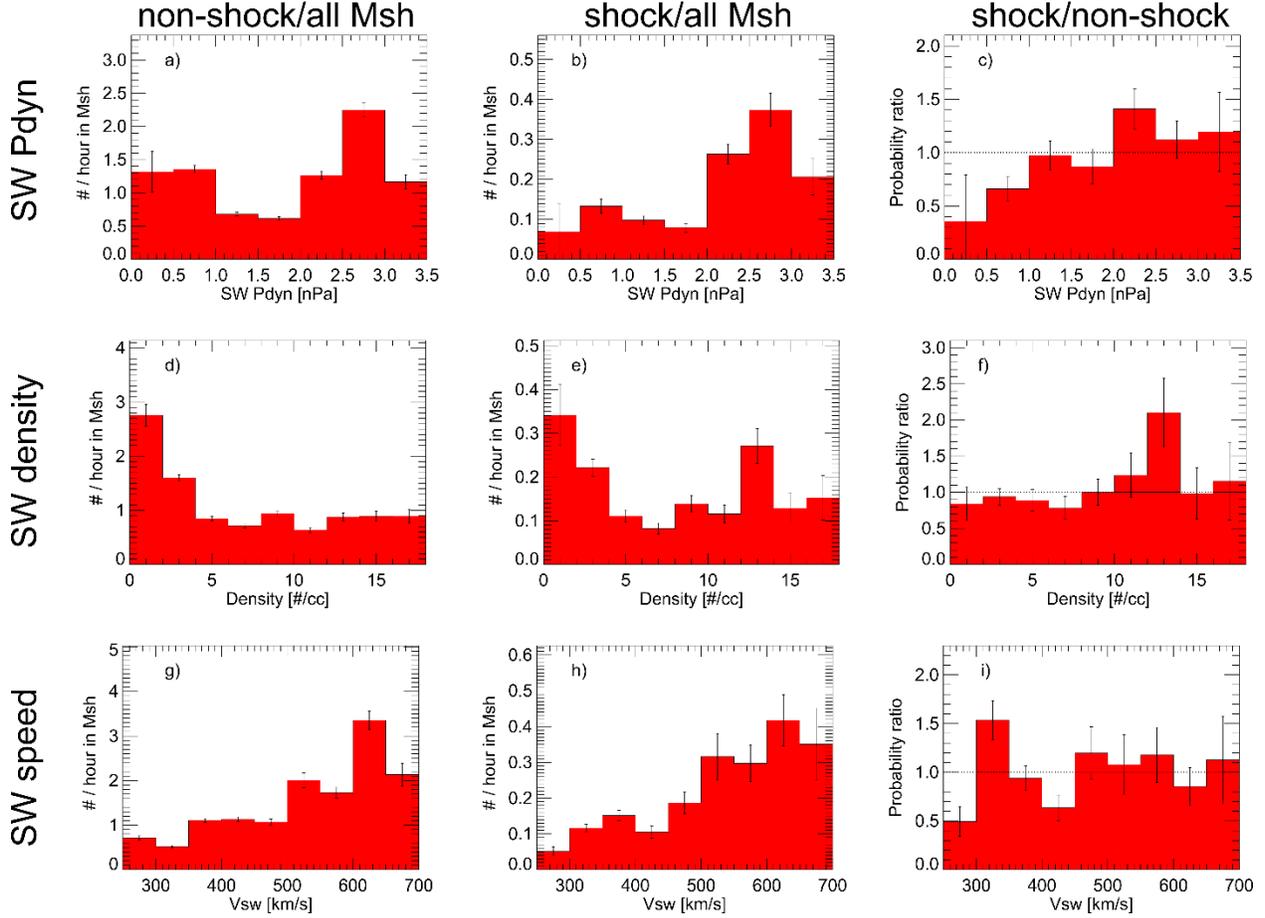

**Figure 2.** (a) The number of non-shock event per hour in the magnetosheath as a function of the solar wind dynamic pressure. (b) The number of shock-like event per hour in the magnetosheath as a function of the solar wind dynamic pressure. (c) The ratio of shock-like event's probability distribution to non-shock event's probability distribution as a function of the solar wind dynamic pressure. (d)-(f) and (g)-(i) are in the same format as (a)-(c) but as a function of the solar wind density and the solar wind speed, respectively. Bin size is equally distributed. We require the bins shown here cover more than 95% of events in total.

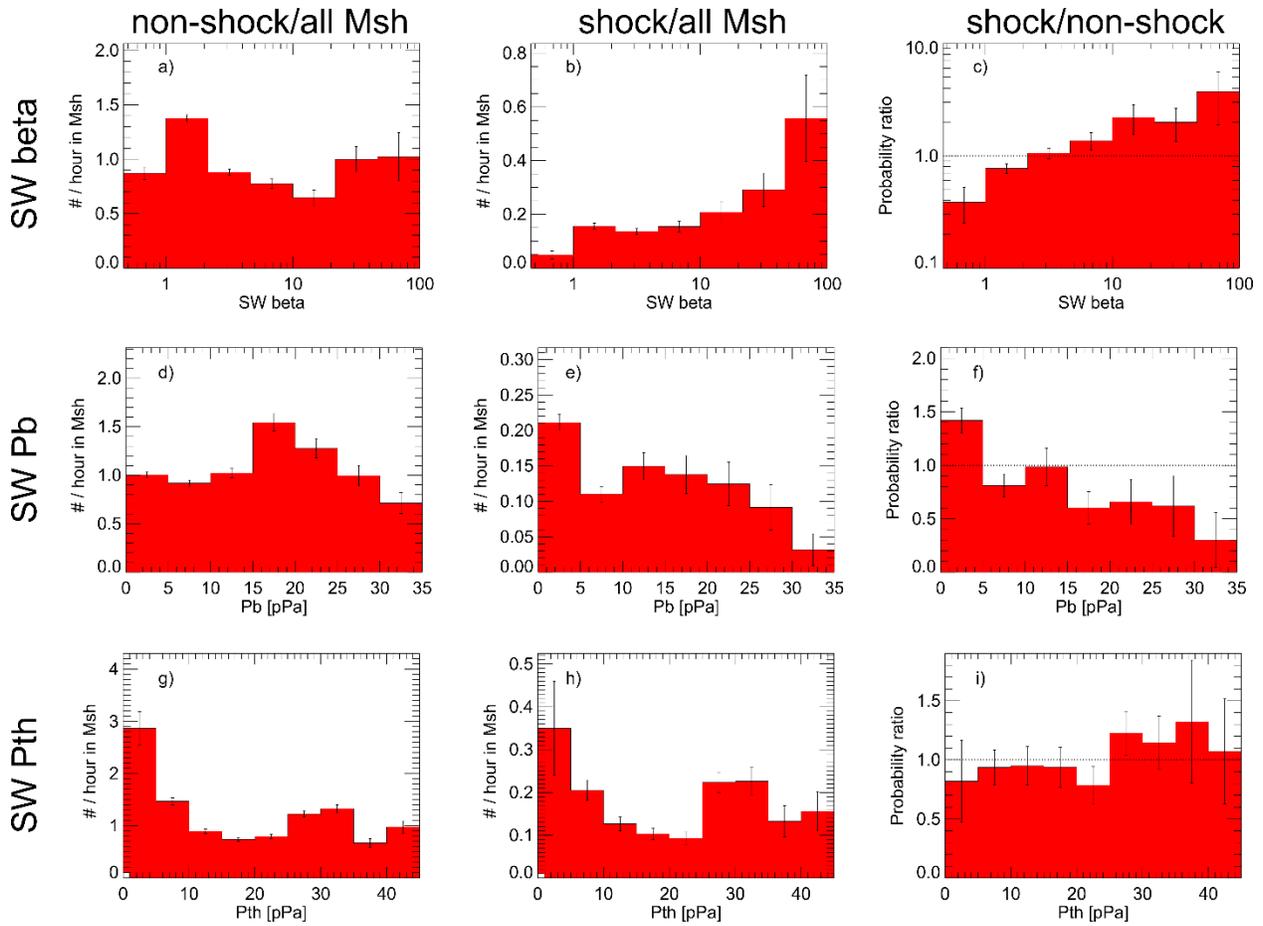

**Figure 3.** Same format as in Figure 2 but as a function of the solar wind plasma beta (top row), the IMF magnetic pressure (middle row), and the solar wind thermal pressure (bottom row; calculated from beta multiplied by the magnetic pressure).

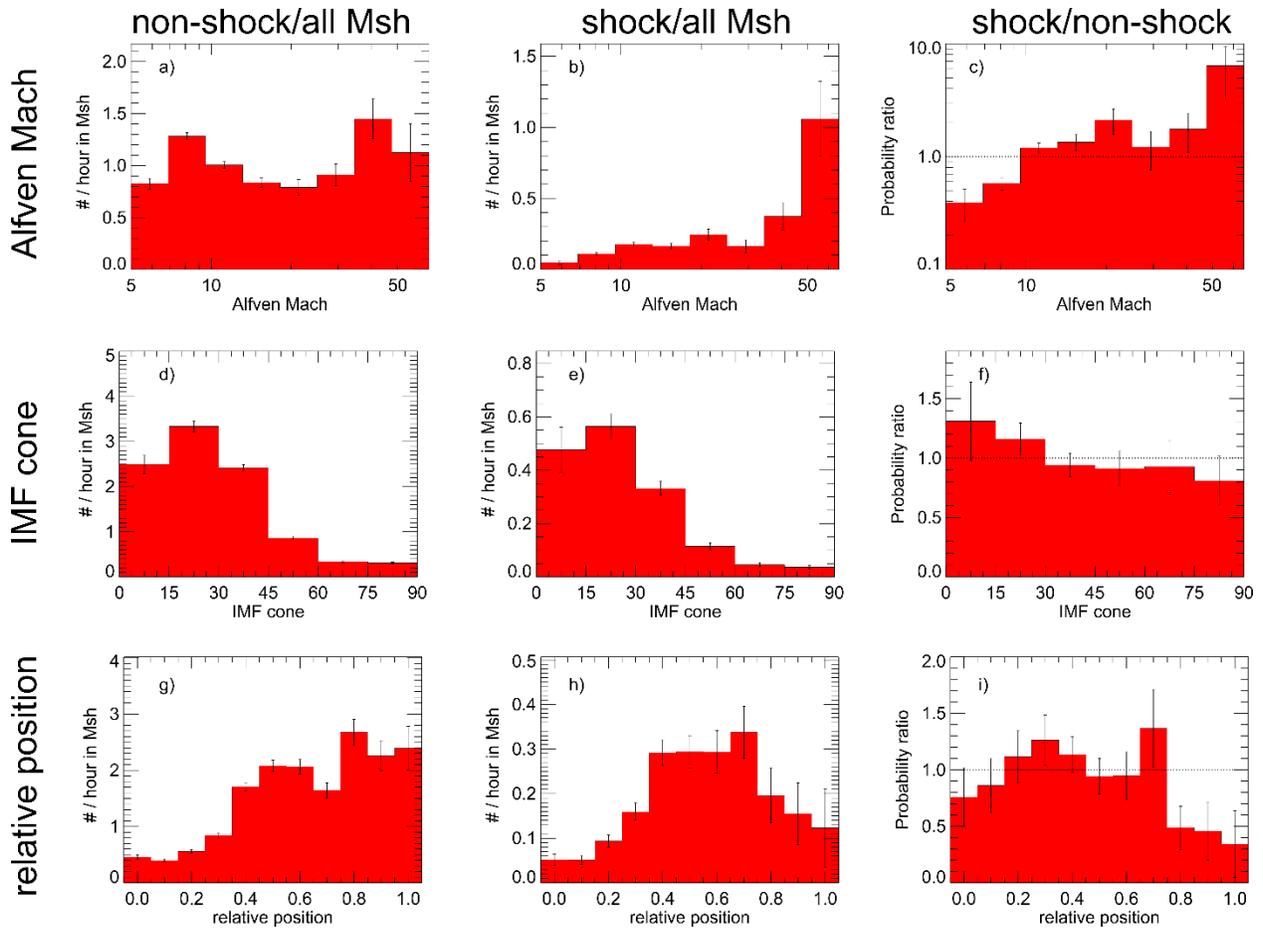

**Figure 4.** Same format as in Figure 2 but as a function of the solar wind Alfvén Mach number (top row), the IMF cone angle (middle row), and the position relative to the bow shock and magnetopause (bottom row; using Merka et al. (2005) model and Shue et al. (1998) model).

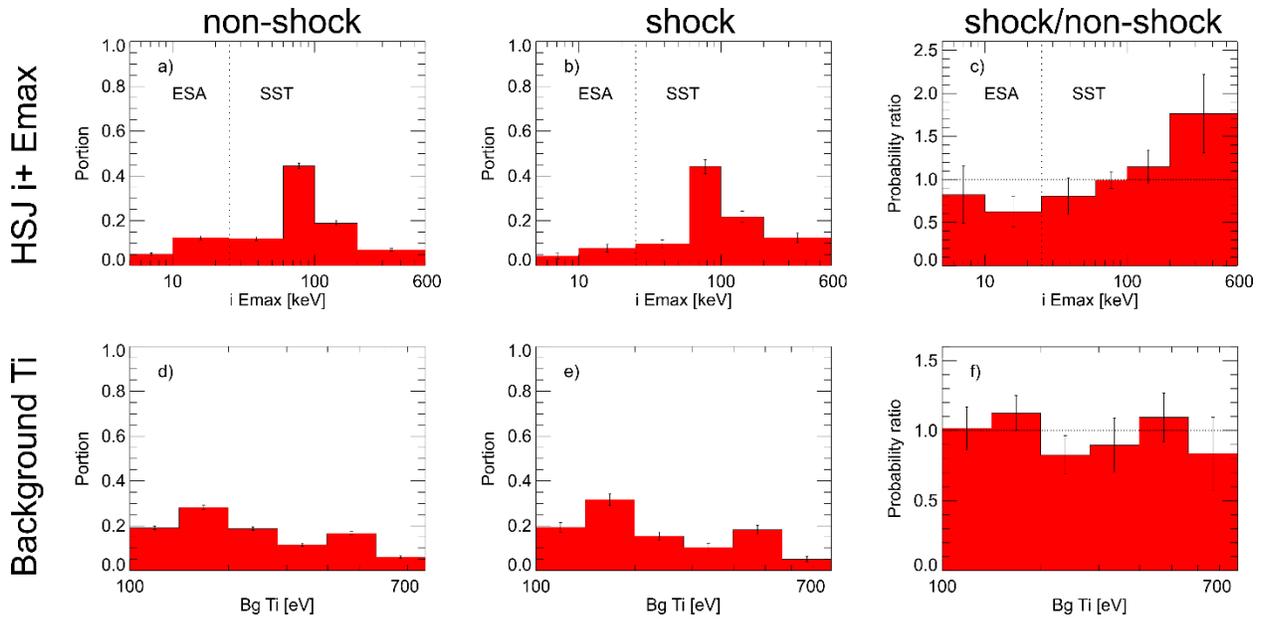

**Figure 5.** (a) The probability distribution of non-shock events as a function of the maximum ion energy. (b) The probability distribution of shock-like events as a function of maximum ion energy. (c) The ratio of (b) to (a). (d)-(f) are in the same format as (a)-(c) but as a function of background magnetosheath ion temperature. HSJ is short for high speed jet.

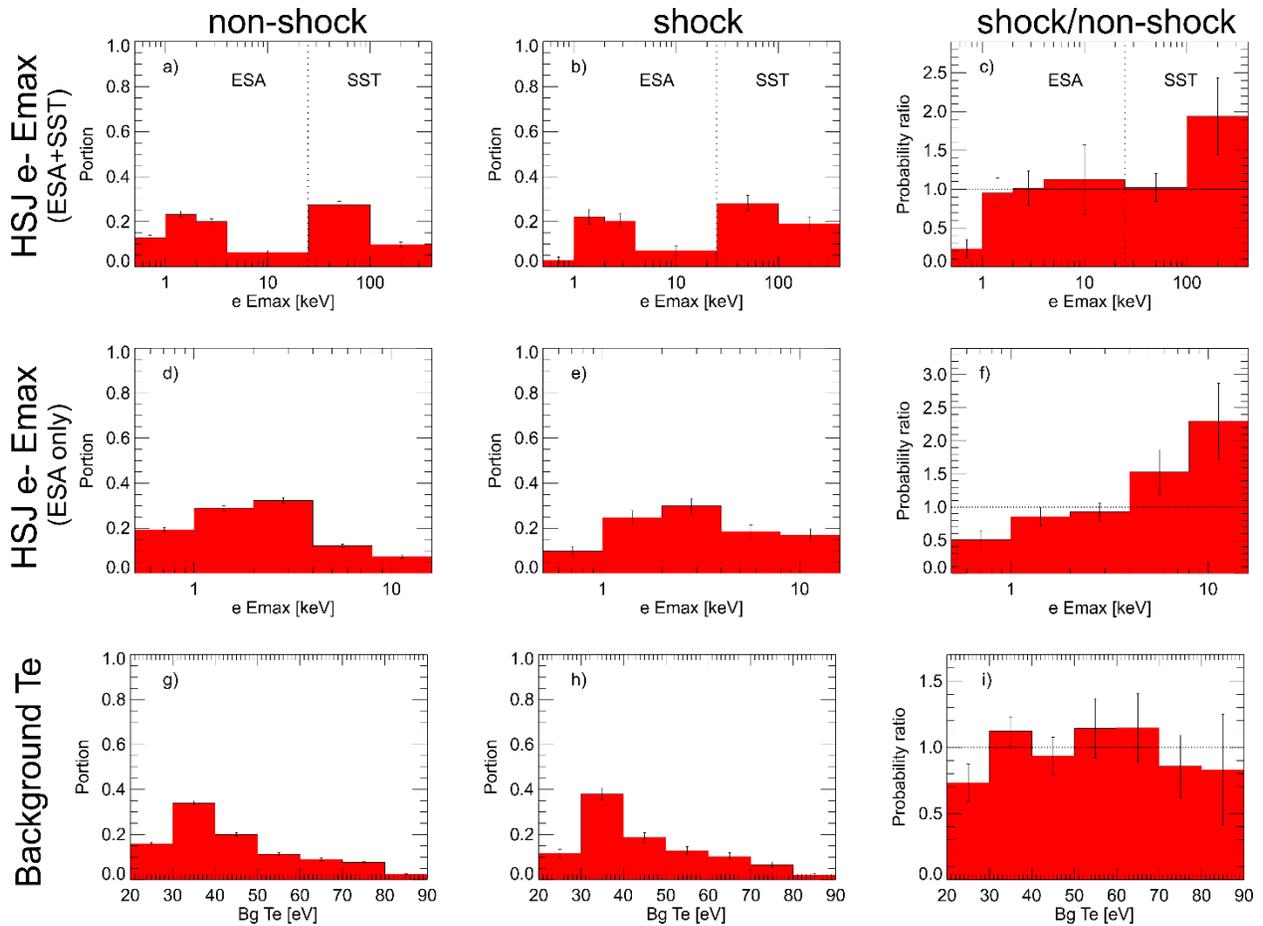

**Figure 6.** Same format as in Figure 5 but as a function of the maximum electron energy measured by ESA and SST (top row) and by ESA only (middle row), and the background magnetosheath electron temperature (bottom row). Note that because SST is more sensitive than ESA, there are very few events at several highest ESA energy channels (10 keV to 25 keV). Thus, we use a larger bin there.

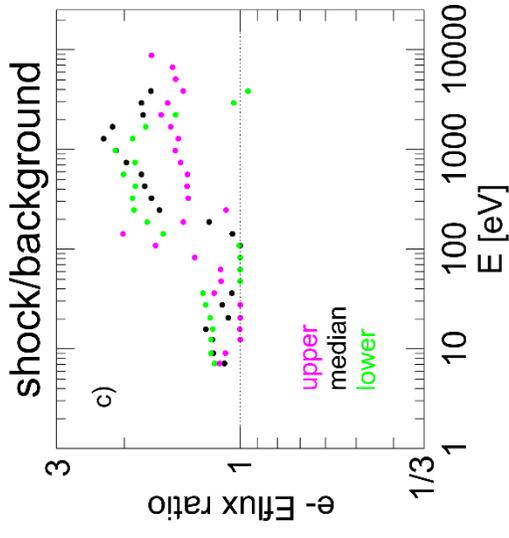
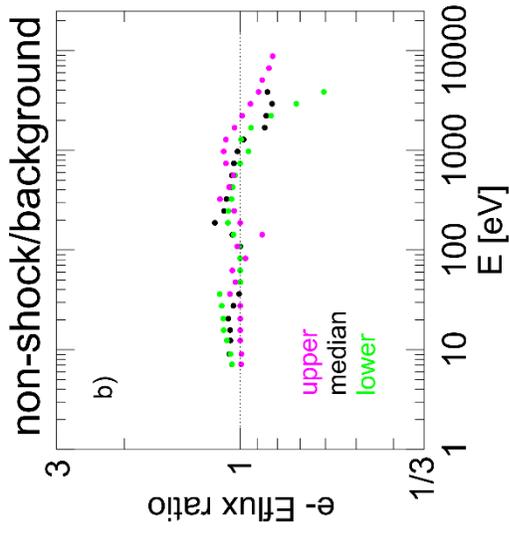
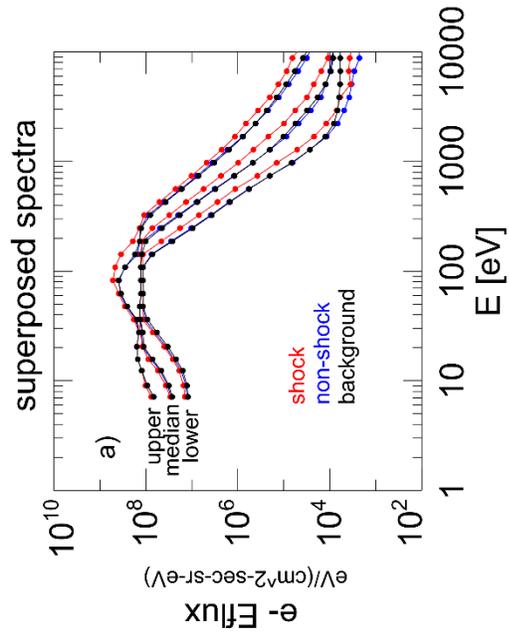

**Figure 7.** (a) The superposed electron energy flux spectra of shock-like events (red), non-shock events (blue), and the background magnetosheath (black) measured by ESA. Three dots at each energy channel are corresponding to the lower quartile (12.5%), median, and upper quartile (78.5%). (b) and (c) are the ratio of electron energy flux values corresponding to median (black), upper quartile (magenta), and lower quartile (green) of superposed non-shock events to the background magnetosheath and superposed shock-like events to the background magnetosheath, respectively. See text for detail. Note that sometimes electron energy flux measurements saturate at $10^8 \, eV/(cm^3 \cdot s \cdot sr \cdot eV)$ in reduced mode.

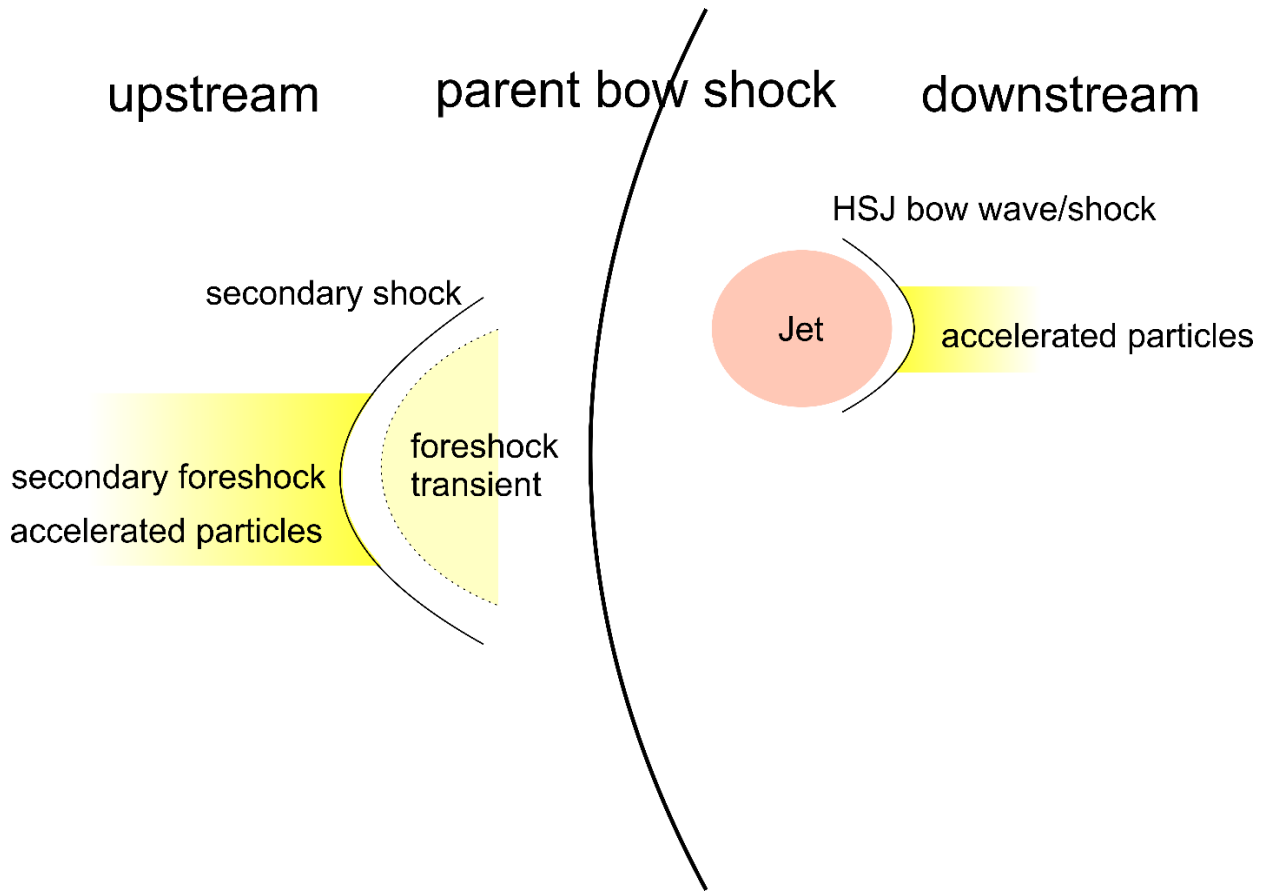

**Figure 8.** A sketch showing that both upstream and downstream of shocks there are nonlinear structures with a secondary bow wave/shock which can accelerate particles. HSJ is short for high speed jet.

**Acknowledgement**

The work at UCLA and SSI was supported by NASA grant NNX17AI45G. TZL is supported by the NASA Living With a Star Jack Eddy Postdoctoral Fellowship Program, administered by the Cooperative Programs for the Advancement of Earth System Science (CPAESS). HH was supported by the Royal Society University Research Fellowship URF\R1\180671 and the Turku Collegium for Science and Medicine. The work in the University of Turku was performed in the framework of the Finnish Centre of Excellence in Research of Sustainable Space. RV acknowledges the financial support of the Academy of Finland (projects 309939 and 312357). We thank the THEMIS software team and NASA's Coordinated Data Analysis Web (CDAWeb, http://cdaweb.gsfc.nasa.gov/) for their analysis tools and data access. OMNI data are available at http://cdaweb.gsfc.nasa.gov/. The THEMIS data and THEMIS software (TDAS, a SPEDAS v3.1 plugin, see Angelopoulos et al. (2019)) are available at http://themis.ssl.berkeley.edu.